# Prediction of the Band Structures of $Bi_2Te_3$-related Binary and Sb/Se-doped Ternary Thermoelectric Materials

Byungki Ryu,[*] Bong-Seo Kim, Ji Eun Lee, Sung-Jae Joo, Bok-Ki Min, HeeWoong Lee, and Sudong Park

*Thermoelectric Conversion Research Center, Korea Electrotechnology Research Institute (KERI), Changwon 51543, Republic of Korea*

Min-Wook Oh,

*Department of Advanced Materials Engineering, Hanbat National University, Daejeon 34158, Republic of Korea*

Density functional calculations are performed to study the band structures of $Bi_2Te_3$-related binary ($Bi_2Te_3$, $Sb_2Te_3$, $Bi_2Se_3$, and $Sb_2Se_3$) and Sb/Se-doped ternary compounds [$(Bi_{1-x}Sb_x)_2Te_3$ and $Bi_2(Te_{1-y}Se_y)_3$]. The band gap was found to be increased by Sb doping and to be monotonically increased by Se doping. In ternary compounds, the change in the conduction band structure is more significant as compared to the change in the valence band structure. The band degeneracy of the valence band maximum is maintained at 6 in binaries and ternaries. However, when going from $Bi_2Te_3$ to $Sb_2Te_3$ ($Bi_2Se_3$), the degeneracy of the conduction band minimum is reduced from 6 to 2(1). Based on the results for the band structures, we suggest suitable stoichiometries of ternary compounds for high thermoelectric performance.

Email: byungkiryu@keri.re.kr

Fax: +82-55-280-1590

**I. INTRODUCTION**

Tetradymite $Bi_2Te_3$ is the traditional thermoelectric (TE) material for direct energy conversion between heat and electricity [1, 2]. For the application of TE technology to waste heat recovery and solid state cooling, it is desired to have high performance *p*- and *n*-type TE materials. The TE energy conversion efficiency and performance are determined by the dimensionless figure of merit ZT defined as $(\alpha^2\sigma/\kappa)T$ [1, 2], where α, σ, κ, and T are the Seebeck coefficient, electrical conductivity, thermal conductivity, and absolute temperature, respectively. Therefore, high TE performance with high ZT can be achieved with a high power factor $\alpha^2\sigma$ (PF) and a small thermal conductivity κ. Many studies have been in the area of low-dimensionalization [3, 4], carrier energy filtering [5, 6] distortion of electronic density of states [7, 8], Fermi level tuning [9-12], band convergence [13, 14], and highly mismatched isoelectric doping [15] to improve the PF. Meanwhile, studies have been done on low total or lattice thermal conductivities in TE materials by using an intrinsically low κ [16, 17], point scattering by alloying [1, 18-20], formation of nano- to micro-structures [21-26], or Peierls distortion and charge density wave instabilities [27].

The electronic band structure is an important property of materials and determines the thermoelectric properties of the Seebeck coefficient α and the electrical conductivity σ. In a simple semi-classical electron transport model with parabolic band and the constant relaxation time approximation, the Seebeck coefficient is given by:

$$\alpha = (2/3e)(k_B\hbar)^2 \, (g\pi/3n)^{2/3} \, (m^*)T, \qquad (1)$$

where *g* is the band valley degeneracy, n is the carrier concentration, and $m^*$ is the effective mass of the single carrier pocket. Note that a high band degeneracy is necessary for high Seebeck coefficient. In addition, the band gap is an important factor and is strongly related to the bipolar effect. When the temperature is increased, the thermal energy becomes comparable to the band gap, and the concentration of minority carrier increases. Thus, the Seebeck coefficient decreases and the thermal conductivity increases.

In a degenerate semiconductor, $|\alpha|$ decreases and $\sigma$ increase when the carrier concentration and the Fermi level increase. Thus, an optimal carrier concentration for maximizing the PF exists. The optimal carrier concentration of $Bi_2Te_3$ is known to be about $10^{19}$ to $10^{20}$ cm$^{-3}$ [10, 12], which corresponds to the heavily-doped system. The anti-site and vacancy defects are revealed to be the major acceptor and donor defects in $Bi_2Te_3$-based materials [28-31].

In $Bi_2Te_3$-based thermoelectric materials, the addition of Sb and Se elements is known to be an effective way to optimize the carrier concentration and to lower the thermal conductivity. Experimentally, the optimal stoichiometry is known to be $(Bi_{1-x}Sb_x)_2Te_3$ (x=0.75) for the *p*-type material and $Bi_2(Te_{1-y}Se_y)_3$ (y=~0.15) for the *n*-type material [1, 2, 30-33]. However, an understanding of why the ternary systems show better thermoelectric behavior then binary compounds is lacking. Although the band structures [34-46] and the thermoelectric transport properties [10, 11, 15, 37, 41-43, 45] have been studied extensively for binary $Bi_2Te_3$, $Sb_2Te_3$, $Bi_2Se_3$, few theoretical studies of band structures and transport properties for $Bi_2Te_3$-based *ternary* compounds have been done [12, 47, 48]. Moreover, minimal information on the band structure, i.e. the band degeneracies and band gaps, is available for ternary systems.

In this study, to understand the effect of Sb and Se alloying on the band structures in $Bi_2Te_3$-based ternary compounds, we performed the density-functional-theory (DFT) electronic structure calculations [49, 50]. We considered various ternary compounds [$(Bi_{1-x}Sb)_2Te_3$ and $Bi_2(Te_{1-y}Se_y)_3$]. The band gaps and the band degeneracies of the band edges were investigated for ternary systems, within virtual crystal approximation (VCA) [51]. Based on the results, we suggest an optimal stoichiometry for $Bi_2Te_3$-based ternary TE compounds.

## II. CALCULATION METHOD

First-principles density functional calculations were performed to study the electronic structure of the $Bi_2Te_3$ based thermoelectric materials. The generalized-gradient-approximation parameterized by Perdue, Burke, and Ernzerhof (PBE) [52] and the projector augmented wave (PAW) pseudopotentials [53] were used, as implemented in Vienna-Ab-initio-Simulation-Package (VASP) code [54, 55]. For Bi, Sb, Te, and Se, the outermost *s* and *p* electrons are considered as valence electrons. The energy cutoff is 250 eV. The experimental lattice parameter and theoretically optimized internal coordinates are used for binary $Bi_2Te_3$, $Sb_2Te_3$, and $Bi_2Se_3$. In the case of $Sb_2Se_3$, the stable structure is the orthorhombic structure. For the comparison with layered $Bi_2Te_3$, the tetradymite $Bi_2Te_3$ atomic structure was used for $Sb_2Se_3$. The lattice parameters were obtained by using the following equations:

$$a[Sb_2Se_3] = a[Sb_2Te_3] + a[Bi_2Se_3] - a[Bi_2Te_3], \qquad (2)$$

and the internal parameter was also relaxed for $Sb_2Se_3$. For ternary alloys of $(Bi_{1-x}Sb_x)_2Te_3$ (**BST**) and $Bi_2(Te_{1-y}Se_y)_3$ (**BTSe**), the lattice parameters and the internal coordinates were interpolated from those for binary compounds. To reduce the heavy computational cost of alloying supercell, we used the primitive unit cell with the VCA. To validate the VCA results, we also tested the (2×2×1) 20-atom alloy supercell for the band-gap calculations and found that the tendency of the band-gap change were maintained. The Monkhorst-Pack *k*-point mesh [56] of gamma-centered 36×36×36 per primitive cell was used. The spin orbit interaction (SOI) was included when calculating the band structures.

## III. RESULTS AND DISCUSSION

The atomic structure of tetradymite-$Bi_2Te_3$ is a layered structure and is categorized in the space group R$\bar{3}$m (No. 166). As shown in the **Figure 1**, two Bi and three Te atoms are stacked with an ABCAB stacking sequence and they form a quintuple layer (QL). The rhombohedral primitive cell contains two inequivalent Te atoms: two Te(1) atoms at the QL surfaces and one Te(2) atom at the center of the QL. The Bi atoms are surrounded by three Te(1) and three Te(2) atoms, forming the local structure of octahedral Bi-Te$_6$. Three QLs are contained in a conventional hexagonal unit cell. **Table 1** shows the details of the atomic structures of Bi$_2$Te$_3$, Sb$_2$Te$_3$, Bi$_2$Se$_3$, and Sb$_2$Se$_3$ binary compounds.

The electronic structure of the Bi$_2$Te$_3$, Sb$_2$Te$_3$, Bi$_2$Se$_3$, and Sb$_2$Se$_3$ binary materials were calculated. Without the spin orbit interaction (SOI), all system had a direct band gap at the Γ *k*-point with band degeneracy (*g*) of 1. As a result, the SOI interaction at Γ was the strongest compared to other *k*-points. The band structures of Bi$_2$Te$_3$, Sb$_2$Te$_3$, Bi$_2$Se$_3$ and Sb$_2$Se$_3$ binary compounds are shown in **Table 1 and Figure 2**. First, we predicted the band gap. For the PBE calculations without SOI, the $E_g^{PBE}$ was the largest for Bi$_2$Te$_3$. With SOI, the PBE gap ($E_g^{SOI}$) decreased to 0.104 eV for Bi$_2$Te$_3$ and increased to 0.111 eV for Sb$_2$Te$_3$. For Bi$_2$Se$_3$, the band gap were nearly maintained. Our result is consistent with those in other DFT reports [39, 44], but still underestimated compared to the experimental values [36, 44, 57, 58] The band gap of Sb$_2$Se$_3$ was found to be 0.201 eV. With the inclusion of SOI, the *k*-points of the valence band maximum (VBM) and conduction band minimum (CBM) ($k_{VBM}$ and $k_{CBM}$) changed and were not on the high symmetry points and lines. As a result, band degeneracies of VBM and CBM ($g_{VBM}$ and $g_{CBM}$) changed. To predict the $k_{VBM}$ and $k_{CBM}$, here we used the 46656 *k*-points (gamma centered 36×36×36) for the 5-atom binary system. Note that, for all binary compounds, the $g_{VBM}$ is always 6, similar to other results [1, 36, 39-41]. However the $g_{CBM}$ is not identical for binaries. For Bi$_2$Te$_3$ and Sb$_2$Se$_3$, the VBM and CBM are at the same *k*-points, showing a direct band gap. In contrary, Sb$_2$Te$_3$ and Sb$_2$Se$_3$ have the indirect band gap. For Sb$_2$Te$_3$, the *k* of CBM is located on the middle of the ΓZ line in Brillouin zone and the band degeneracy is only 2. In the case of Bi$_2$Se$_3$, the CBM is at Γ. We'd like to mention that the $g_{CBM}$ can be sensitive to the choice of pseudopotentials or exchange-correlation energies and can vary from 6 to 2(1) due to the small energy difference between local minima [35, 40]. Note that our density functional calculation result of $g_{CBM} = 6$ for Bi$_2$Te$_3$ is consistent to that from recent quasiparticle GW approximation calculations [40].

The electronic structures of (Bi$_{1-x}$Sb$_x$)$_2$Te$_3$ (x = 0 to 1) and Bi$_2$(Te$_{1-y}$Se$_y$)$_3$ (y = 0 to 1) alloys were investigated. All lattice parameters and internal parameters were linearly interpolated by using the values for binaries. We adopted the VCA when calculating the electronic structures. The disordered ternary systems were considered as ordered binary systems with interpolated atomic potentials, ignoring any possible short range ordering. **Figure 3** shows the $E_g^{SOI}$ for $(Bi_{1-x}Sb_x)_2Te_3$ and $Bi_2(Te_{1-y}Se_y)_3$ ternary alloy compounds for gamma-centered $k$-point meshes of 6×6×6, 12×12×12, 24×24×24, and 36×36×36. Note that the very fine $k$-point mesh is mandatory to ensure the convergence of the band gap for $Bi_2Te_3$-based ternary compounds as well as for binary compounds. The band-energy convergence with respect to the $k$-point mesh were calculated to be less than 1 meV. We found that the band gap of $E_g^{SOI}$ was increased by alloying with Sb and Se elements. The $E_g^{SOI}$ of $Sb_2Te_3$ was larger than that of $Bi_2Te_3$ by 0.012 eV. The band gap of **BST** increased with increasing x and became maximum at x = 30–70%. On the other hand, the band gap of **BTSe** monotonically increased from 0.105 to 0.260 eV when going from $Bi_2Te_3$ to $Bi_2Se_3$.

By comparing the VCA band gap with the supercell band gaps, we verified the reliability of the band gap of the primitive cell with the VCA. For BST and BTSe, we used the (2×2×1) 20-atom supercell for BST and BTS that contained 8 Bi/Sb atoms and 12 Te/Se atoms. We considered 12 different atomic configurations for the $Bi_{8-n}Sb_nTe_{12}$ supercells ($n$ is an integer from 0 to 8) and 25 configurations for the $Bi_8Te_{12-m}Se_m$ ($m$ is an integer from 0 to 12) supercells. In the modelled structures, the lattice parameters and the internal parameters were interpolated from those for binary structures. The values of $E_g^{SOI}$ from the VCA and the supercell approaches are compared in **Figure 4**. The supercell band gap was calculated to be smaller than the VCA band gap. We think that the disorder-induced charge-density-localization in a finite supercell might be responsible for the band-gap variation in the supercell calculations, as compared to the VCA results. For that reason, we calculated and used the fitting curves for the BST and the BTSe band gaps. For **BST**, the fitted band-gap curve, represented by dashed line, is nearly the same as that for supercell approach. However, for **BTSe**, the tendency of the fitted band gap change is still valid.

Next, the positions of the VBM and the CBM were examined for ternary compounds. **Figure 5** represents the positions of the VBM and the CBM, $k_{VBM}$ and $k_{CBM}$, for the **BST** and the **BTSe** alloys. We found that the change in $k_{CBM}$ was significant when alloying while the change in $k_{VBM}$ was relatively small due to the valence band shapes being similar among the binary compounds. For small x and y, the **BST** and the **BTSe** band edges are located near the $k_{VBM}$ and the $k_{CBM}$ of $Bi_2Te_3$. However, as x and y were increased, the band edges moved toward the $k_{VBM}$ and the $k_{CBM}$ of $Sb_2Te_3$ and $Bi_2Se_3$, respectively. For $Bi_2Te_3$, the band degeneracy of the VBM and the CBM was 6. We found that the band degeneracy $g_{VBM}$ = 6 is robust against the $Bi_2Te_3$ alloying with $Sb_2Te_3$ and $Bi_2Se_3$. To the contrary, $g_{CBM}$ changed when alloying. For BST, the $g_{CBM}$ was 6 for x = 0.0 – 0.2 and the $g_{CBM}$ was 2 for x = 0.3 – 1.0. For BTS, the $g_{CBM}$ was 6 for y = 0.0 – 0.3, the $g_{CBM}$ was 2 for y = 0.4 – 0.6, and the $g_{CBM}$ = 1 when y = 0.7 – 1.0.

For high thermoelectric performance, a large power factor and a small thermal conductivity are needed. The power factor is maximized with optimal band structures with high band degeneracy, relatively large band gap, and proper position of the Fermi level. Based on calculated results, band gaps and band degeneracies of $Bi_2Te_3$-based ternary alloys, we suggest a suitable stoichiometry for high thermoelectric performance. For *p*-type **BST**, the band degeneracy is maintained for all stoichiometry. If our VCA band gap is correct, $(Bi_{1-x}Sb)_2Te_3$ with x = 0.3 – 0.7 seems to be good, which might be responsible for the smaller bipolar effect on Seebeck coeffieint and the thermal conductivity. However, if the supercell band gap is correct, there should be a band gap reduction and the increase of bipolar effect in Seebeck coefficient and the electrical thermal conductivity. In this case, the role of Sb-doping/alloying in $Bi_2Te_3$ is to generate the *p*-type carriers by the formation of anti-site defects and a reduction in the lattice thermal conductivity caused by point disorder. For *n*-type **BTSe**, $Bi_2(Te_{1-y}Se_y)_3$ with y = 0.3 is the best due to its having the largest band gap among the high-CBM-degeneracy alloys ($g_{CBM}$ = 6). We find that our suggested stoichiometries for **BST** and **BTSe** (x = 0.3 – 0.7, y = 0.3) are consistent with the stoichiometries used in experiments (x = 0.75, y = ~0.15) [1, 2, 30, 33]. Note that here we only consider the band gaps and the band degeneracy and do not consider the reduction in the lattice thermal conductivity in disordered alloy ternary compounds. If we consider the reduction in the lattice thermal conductivity caused by point disorder in ternary compounds, we may conclude that the thermoelectric property of ternary compounds is superior to that of binary compounds.

## IV. CONCLUSION

In conclusion, we investigated the electronic structures of $Bi_2Te_3$-based ternary alloys within density functional theory and the virtual crystal approximation. The band degeneracy, the band gap, and the positions of the band edges of $(Bi_{1-x}Sb_x)_2Te_3$ and $Bi_2(Te_{1-y}Se_y)_3$ were systemically examined. The band gap was found to be increased by Sb doping and to be monotonically increased by Se doping. For *p*-type **BST**, the valence band degeneracy were maintained. However, for *n*-type **BTSe**, the conduction band degeneracy was very sensitive to the alloy stoichiometry. Based on the results, we suggest a suitable stoichiometry of $Bi_2Te_3$-based ternary alloys for better thermoelectric performance.

## ACKNOWLEDGMENTS

This research was supported by Korea Electrotechnology Research Institute (KERI) Primary research program through the National Research Council of Science & Technology (NST) funded by the Ministry of Science, ICT and Future Planning (MSIP) [No. 15-12-N0101-24, *Development of design tools of thermoelectric and energy materials*] and by the IT R&D program of Korea Evaluation Institute of Industrial Technology (KEIT) funded by Ministry of Trade, Industry and Energy (MOTIE) [No. 10048035, *Development of Bi-Te based Thermoelectric Energy Conversion Materials by Controlling Multi-level Nanostructure in Pilot Scale*].

Table 1. Atomic and electronic structure table of binary $Bi_2Te_3$, $Sb_2Te_3$, $Bi_2Se_3$, and $Sb_2Se_3$ compounds. *a* and *c* are the lattice constants of the conventional hexagonal unit cell, where $\|a_{RHL}\|$ is the length of the rhombohedral lattice vector and *cos* α is the directional cosine between the rhombohderal lattice vectors. *u* and *v* are internal coordinates for Bi and Te(1). $E_g^{SOI}$ and $E_g^{PBE}$ are the PBE band gap with and without spin-orbit-interaction (SOI). The band degeneracies (*g*) of the VBM and the CBM are denoted by $g_{VBM}$ and $g_{CBM}$. The type of band gap (type of $E_g$) is classified as the direct and indirect gaps. The *k*-points of the VBM and the CBM are denoted as $k_{VBM}$ and $k_{CBM}$. All *k* vectors are represented by using the basis of the reciprocal lattices.

|  | $Bi_2Te_3$ | $Sb_2Te_3$ | $Bi_2Se_3$ | $Sb_2Se_3$ |
|---|---|---|---|---|
| *a* (Å) | 4.383 | 4.250 | 4.138 | 4.004* |
| *c* (Å) | 30.487 | 30.400 | 28.640 | 28.553* |
| $\|a_{RHL}\|$ (Å) | 10.473 | 10.426 | 9.841 | 9.794 |
| *cos* α | 0.912 | 0.917 | 0.912 | 0.916 |
| *u* | 0.400 | 0.399 | 0.401 | 0.399 |
| *v* | 0.210 | 0.211 | 0.211 | 0.212 |
| $E_g^{PBE}$ (eV) | 0.271 | 0.027 | 0.185 | 0.174 |
| $E_g^{SOI}$ (eV) | 0.105 | 0.133 | 0.260 | 0.201 |
| $g_{VBM}$ | 6 | 6 | 6 | 6 |
| $g_{CBM}$ | 6 | 2 | 1 | 6 |
| Type of $E_g$ | Direct | Indirect | Indirect | Direct |
| $k_{VBM}$ | (.417, .417, .333) | (.389, .389, .333) | (.361, .361, .278) | (.028, .028, .028) |
| $k_{CBM}$ | (.417, .417, .333) | (.194, .194, .194) | (0, 0, 0) | (.028, .028, .028) |

Figure Captions.

**Fig. 1.** Ball and stick model of the atomic structure of $Bi_2Te_3$. Bi/Sb atoms are denoted by large balls, and Te/Se atoms are denoted by smaller ones. The Bi/Sb atoms are surrounded by six Te/Se atoms: three Te/Se(1) at QL surfaces and three Te/Se(2) at the center of the QL.

**Fig. 2.** Band structures are drawn for (a) $Bi_2Te_3$, (b) $Sb_2Te_3$, (c) $Bi_2Se_3$, and (4) $Sb_2Se_3$ binary compounds in a tetradymite rhombohedral structure. The $E_{VBM}$ is set to zero. The dashed vertical lines indicate $E_{CBM}$ and $E_{VBM}$. The $k$-point **a** corresponds to (0.43, 0.43, 0.64) on the basis of the reciprocal lattices.

**Fig. 3.** The $E_g^{SOI}$ for (a) $(Bi_{1-x}Sb_x)_2Te_3$ and (b) $Bi_2(Te_{1-y}Se_y)_3$ ternary alloy compounds for gamma centered $k$-point meshes of 6×6×6, 12×12×12, 24×24×24, and 36×36×36. The dashed horizontal lines represent the band gaps of the binary compounds.

**Fig. 4.** The $E_g^{SOI}$ for (a) $(Bi_{1-x}Sb_x)_2Te_3$ and (b) $Bi_2(Te_{1-y}Se_y)_3$ obtained by using the VCA approach (filled circles) and the supercell approach (open circles). The solid and the dashed lines are lines fitted to the band gaps obtained by using the VCA and the supercell approach, respectively.

**Fig. 5.** The positions of band edges are mapped for (a) $(Bi_{1-x}Sb_x)_2Te_3$ and (b) $Bi_2(Te_{1-y}Se_y)_3$. $b_1$, $b_2$, and $b_3$ are the reciprocal vectors for the given rhombohedral system. $k_{VBM}$ and $k_{CBM}$ are represented by filled circles and X marks, respectively. The numbers in the figures are the compositional ratios of the BST and the BTS alloys. For BST and BTS, the movement of $k_{CBM}$ is significant as compared to that of $k_{VBM}$.

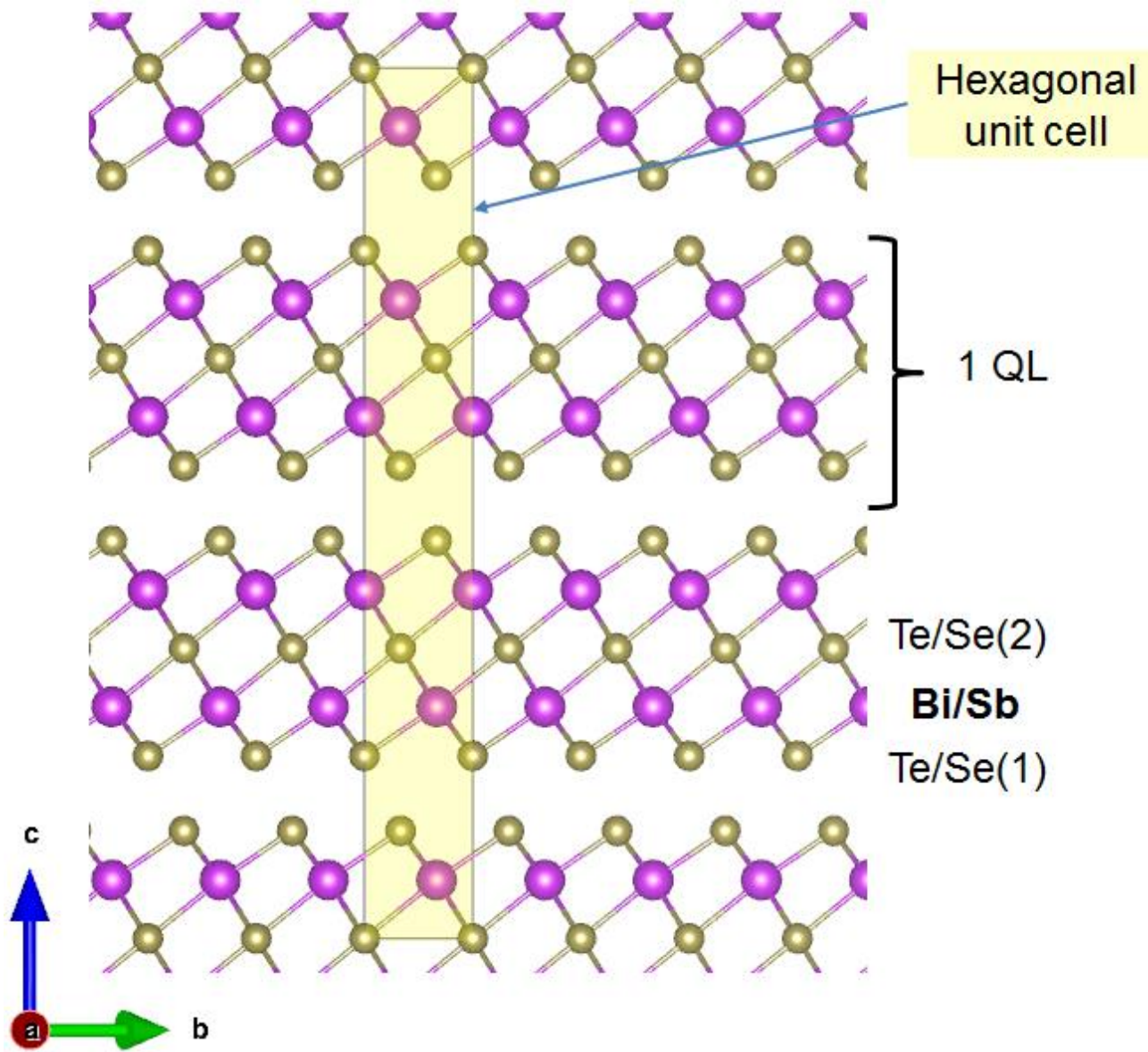

Figure 1

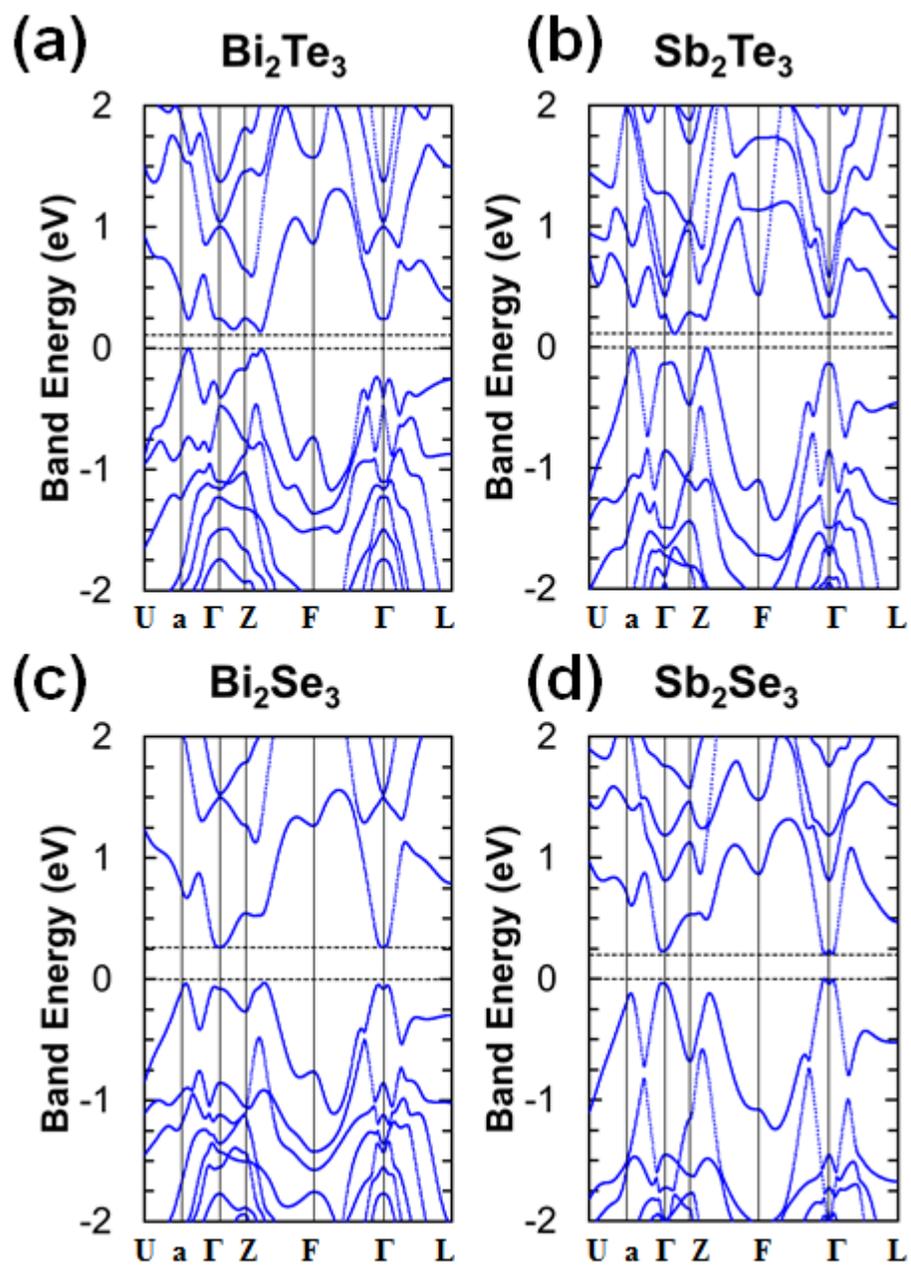

Figure 2

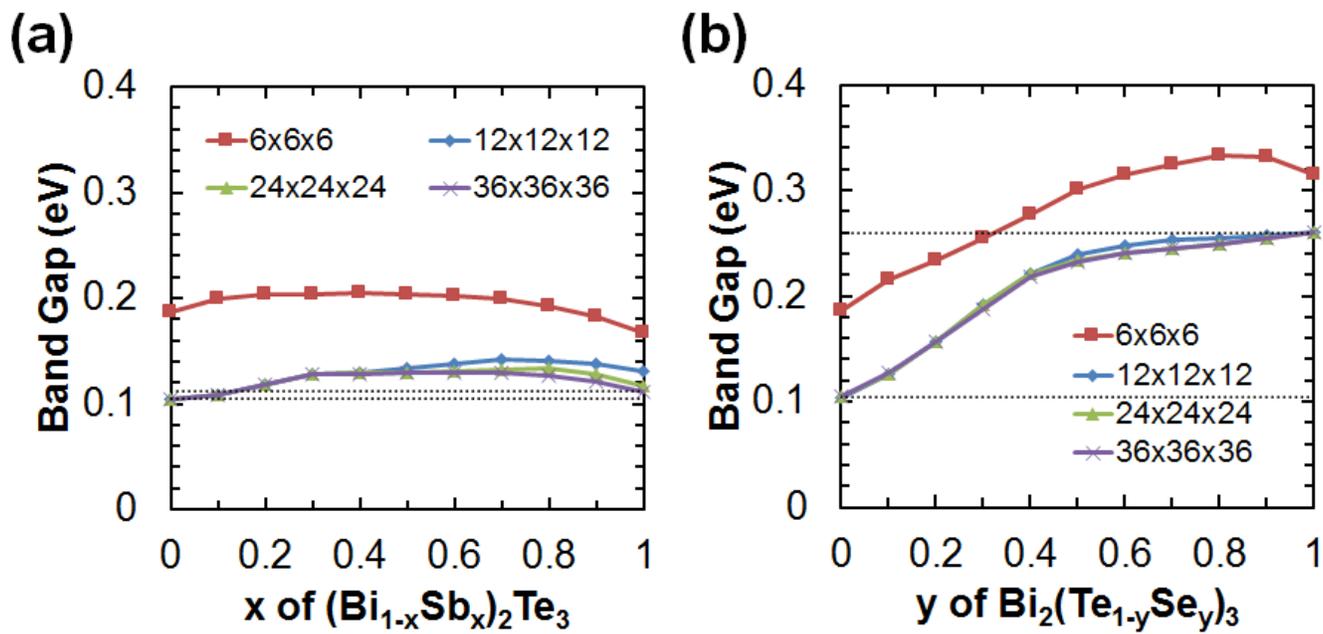

Figure 3

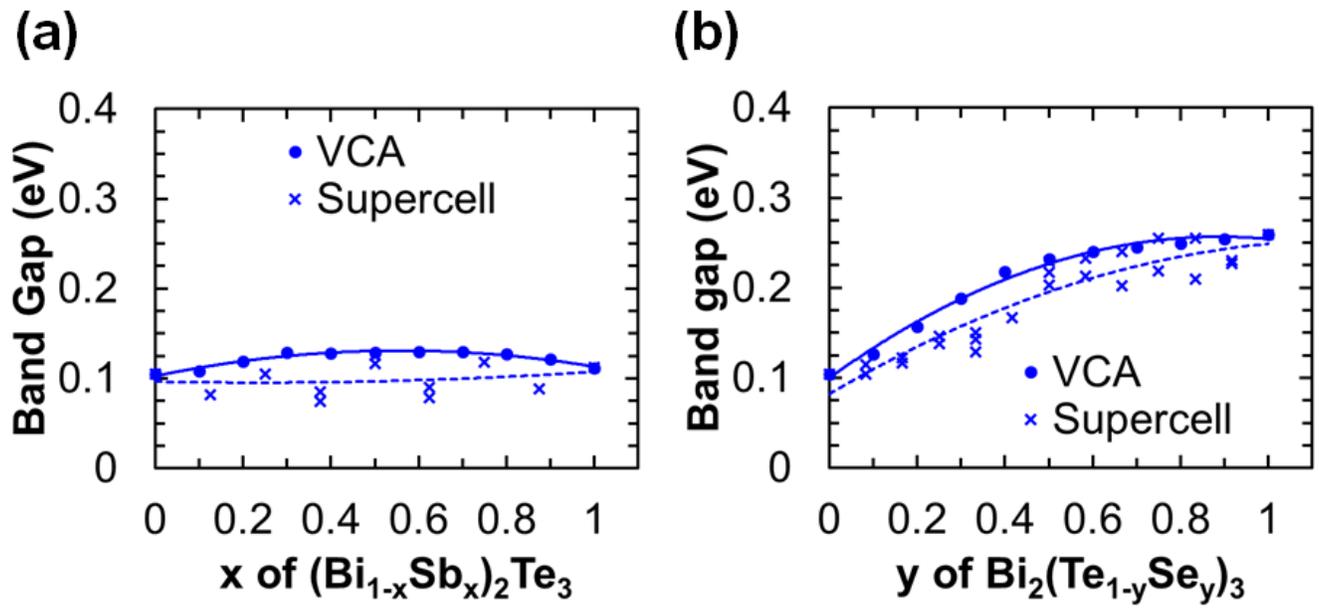

Figure 4

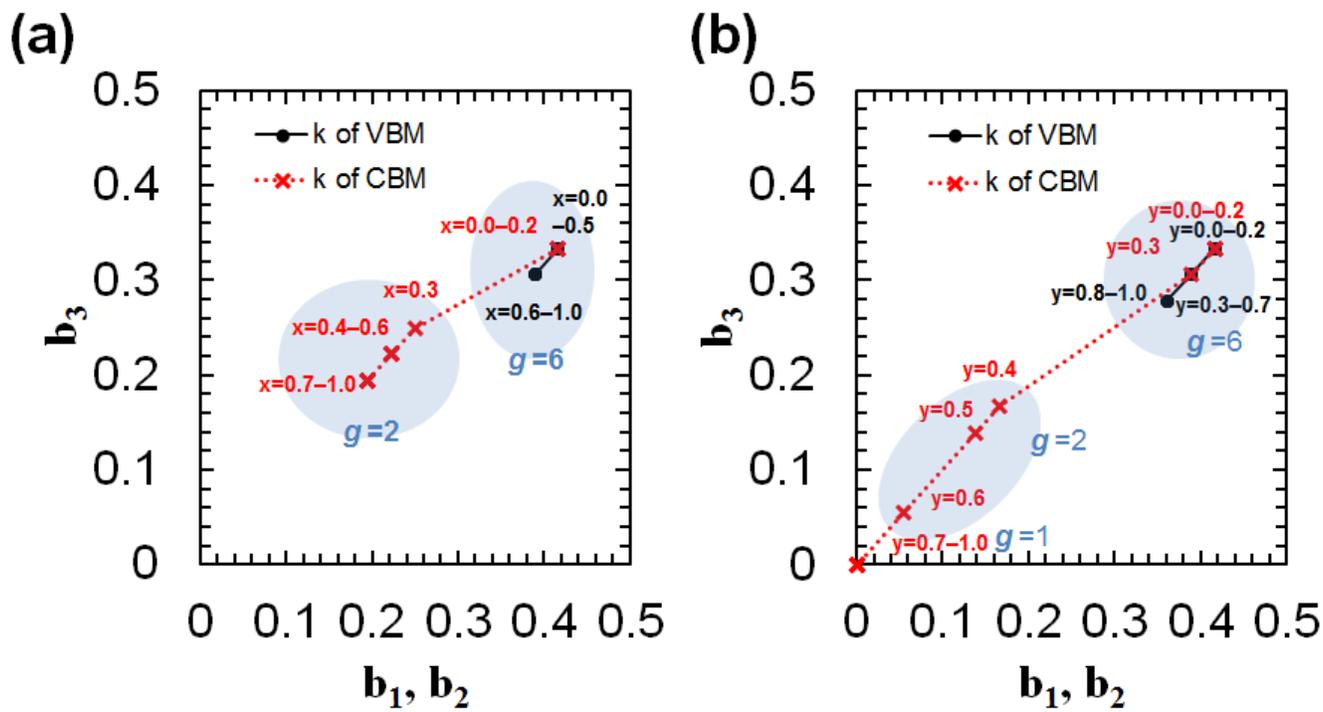

Figure 5